\documentstyle[epsfig,longtable]{aipproc}

\begin{document}
\title{The Phases of QCD in\\
Heavy Ion Collisions and Compact Stars}

\author{Krishna Rajagopal\thanks{One version
of this review is to appear in the Comments on Nuclear
and Particle Physics section of 
Comments on Modern Physics. Other versions were
contributed to the Proceedings of the Conference
on Intersections of Nuclear and Particle Physics, Qu\'ebec, May 2000
and to the Proceedings of the 40th Zakopane School of
Theoretical Physics, June 2000.
Many thanks 
to the CIPANP organizers for a conference which was stimulating precisely
because 
it addressed so many facets of the
intersection between nuclear and particle physics. Many
thanks to
the Zakopane organizers for a school which brought condensed
matter physics and physicists and particle physics 
and physicists together, and for an excuse to visit a beautiful
part of the world for the first time. 
I am grateful to 
M.~Alford, B.~Berdnikov,
J.~Berges, J.~Bowers, E.~Shuster, E.~Shuryak, M.~Stephanov and F.~Wilczek
for fruitful collaboration. I acknowledge helpful
discussions with P.~Bedaque, D.~Blaschke,
I.~Bombaci, G. Carter,
D. Chakrabarty,
J. Madsen, C. Nayak, M. Prakash, 
D.~Psaltis, S. Reddy, M.~Ruderman, T. Sch\"afer, A. Sedrakian,
D. Son, I. Wasserman and F. Weber.
This work is supported in 
part  by the U.S. Department
of Energy (D.O.E.) under cooperative research agreement \#DF-FC02-94ER40818
and by a DOE OJI Award and by the
Alfred P. Sloan Foundation. Preprint MIT-CTP-3020.} }
\address{Center for Theoretical Physics, Massachusetts Institute of 
Technology\\
Cambridge, MA 02139}

\maketitle

\begin{abstract}
I review arguments for the existence of a critical point $E$
in the QCD phase diagram as a function of temperature $T$ and
baryon chemical potential $\mu$. I describe how heavy ion collision
experiments at the SPS and RHIC can discover the tell-tale
signatures of such a critical point, thus mapping this
region of the QCD phase diagram.  I then review the
phenomena expected in cold dense quark matter: color superconductivity
and color-flavor locking. I close with a snapshot of ongoing
explorations of the implications of recent developments
in our understanding of cold dense quark matter for 
the physics of compact stars.
\end{abstract}

The QCD vacuum in which we live, which has the familiar hadrons
as its excitations, is but one phase of QCD, and far from
the simplest one at that. One way to better understand
this phase and the nonperturbative dynamics of QCD more generally
is to study other phases and the transitions between phases.
We are engaged in a voyage
of exploration, mapping the QCD phase diagram as a function
of temperature $T$ and baryon number chemical potential $\mu$.
Because QCD is asymptotically free, its high temperature and
high baryon density phases are more simply and more
appropriately described in terms of quarks and gluons as
degrees of freedom, rather than hadrons. The chiral symmetry
breaking condensate which characterizes the vacuum phase
melts away.  At high temperatures, in the resulting quark-gluon plasma
(QGP) phase all of the symmetries of the QCD Lagrangian are unbroken
and the excitations have the quantum numbers of quarks and gluons.
At high densities, on the other hand,
quarks form Cooper pairs and new
condensates develop.  The formation
of such superconducting
phases~\cite{Barrois,BailinLove,ARW1,RappETC,CFL} 
requires only weak attractive interactions; these phases may nevertheless
break chiral symmetry~\cite{CFL} 
and have excitations with the same quantum numbers as
those in a confined phase~\cite{CFL,SW1,ABR2+1,SW2}.   These 
cold dense quark matter phases may arise in the core of
neutron stars; mapping this region of the phase diagram
requires an interplay between theory and neutron star
phenomenology.  We describe efforts in this direction in Section IV. 
A central goal of the experimental heavy ion
physics program is to explore and map the higher temperature
regions of the QCD phase diagram. 
Recent theoretical developments
suggest that a key qualitative feature, namely a critical
point which in a sense defines the landscape
to be mapped, may be within reach of discovery and analysis  
as data is taken at several different
energies~\cite{SRS1,SRS2}. The discovery of the critical point
would transform this region of
the map of the QCD phase
diagram from one based only on
reasonable inference from universality, lattice gauge theory
and models into one with a solid experimental basis.

\section{The Critical Point}

We begin our walk through the phase diagram at zero
baryon number density, with a brief review~\cite{rajreview} 
of the phase changes
which occur as a function of temperature.
That is, we begin by restricting ourselves to 
the vertical axis in Figures~1 through 4.  This slice of
the phase diagram was explored by the early universe
during the first tens of microseconds after the big bang
and can be studied in lattice simulations.  As heavy ion collisions
are performed at higher and higher energies, they create plasmas
with a lower and lower baryon number to entropy ratio and therefore
explore regions of the phase diagram 
closer and closer to the vertical axis.

\begin{figure}[t]
\begin{center}
\vspace{-0.1in}
\hspace*{0in}
\epsfysize=2.5in
\epsffile{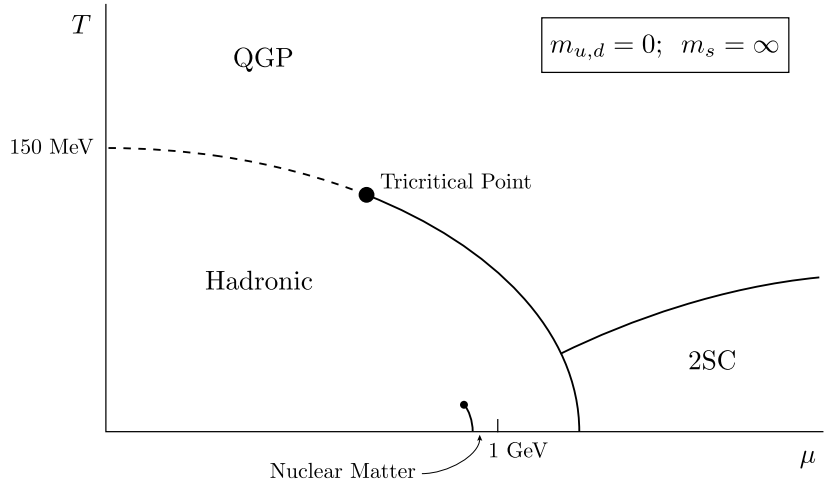}
\end{center} \label{fig1}
\caption{QCD Phase diagram for two massless quarks. Chiral symmetry
is broken in the hadronic phase and is restored elsewhere in the 
diagram. The chiral phase transition changes from second to
first order at a tricritical point. The phase at high density and low
temperature is a color superconductor in which up and down
quarks with two out of three colors pair and form a condensate.
The transition between this 2SC phase and the QGP phase 
is likely first order. The transition on the horizontal
axis between the hadronic and 2SC phases is first 
order. The transition between a nuclear matter ``liquid''
and a gas of individual nucleons is also marked; it ends
at a critical point at a 
temperature of order 10~MeV, characteristic of the forces
which bind nucleons into nuclei.}
\vspace{-0.1in}
\end{figure}
In QCD with
two massless quarks ($m_{u,d}=0$; $m_s=\infty$; Figure~1)
the vacuum phase, with hadrons as excitations, is characterized
by a chiral condensate $\langle \bar\psi_{L\alpha a}\psi_R^{\alpha a}\rangle$.
(The color index $\alpha$ is summed over the three colors; the 
flavor index $a$ is summed over the two flavors.) Whereas
the QCD Lagrangian is invariant under separate global flavor
rotations of the left-handed and right-handed
quarks, the presence of the chiral condensate spontaneously
breaks $SU(2)_L \times SU(2)_R$ to the subgroup $SU(2)_{L+R}$,
in which only  simultaneous flavor rotations of $L$ and $R$ quarks
are allowed.  In this way, locking left- and right-handed rotations 
breaks global symmetries and 
results in three massless Goldstone bosons, the pions.
The chiral order parameter, a $2\times 2$ matrix $M^{ab}$ in
flavor space,
can be written in terms of
four real fields $\sigma$ and $\vec\pi$ as
\begin{equation}
\langle \bar\psi_{L\alpha}^a\psi_{R}^{\alpha b}\rangle = M^{ab} = 
\sigma \delta^{ab} + \vec \pi \cdot \left(\vec\tau\right)^{ab} \ ,
\end{equation}
where the $\vec\tau$ are the three Pauli matrices.
$SU(2)_L$ and $SU(2)_R$ rotations act on $M^{ab}$ from the
left and right, respectively.  
The order parameter
can also be written as a four component scalar 
field $\phi =(\sigma,\vec\pi)$
and the  $SU(2)_L \times SU(2)_R$ rotations are then simply
$O(4)$ rotations of $\phi$.  In this language, the symmetry breaking
pattern $SU(2)_L \times SU(2)_R\rightarrow SU(2)_{L+R}$ is
described as
$O(4)\rightarrow O(3)$: in the vacuum, $\langle\phi\rangle\neq 0$ 
and this condensate picks a direction in $O(4)$-space.
The direction in which the condensate points is
conventionally taken
to be the $\sigma$ direction.  In the presence 
of $\langle\sigma\rangle\neq 0$, the $\vec\pi$ excitations
are excitations of the direction in which $\langle\phi\rangle$
is pointing, and are therefore
massless goldstone modes.

At nonzero but low temperature, one finds a gas of pions,
the analogue of a gas of spin-waves, but $\langle\phi\rangle$
is still nonzero.  Above some temperature $T_c$, entropy
wins over order (the direction in which $\phi$
points is scrambled) and $\langle\phi\rangle=0$.
The phase transition at which chiral symmetry is restored 
is likely second order and belongs to the universality
class of $O(4)$ spin models in three dimensions~\cite{piswil}.
Below $T_c$, chiral symmetry is broken and there are three
massless pions.  At $T=T_c$, there are four massless degrees
of freedom: the pions and the sigma. Above $T=T_c$, the pion
and sigma correlation lengths are degenerate and finite.

In nature, the light quarks are not massless.  Because
of this explicit chiral symmetry breaking,
the second order phase transition is replaced by an 
analytical crossover: physics changes dramatically but smoothly in the 
crossover region, and no correlation length diverges.
Thus, in Figure~2, there is no sharp boundary on the 
vertical axis separating the low temperature hadronic world
from the high temperature quark-gluon plasma.  
This picture is consistent with present lattice 
simulations~\cite{latticereview,latestlattice},
which suggest $T_c\sim 140-190$~MeV~\cite{latticeTc,latestlattice}.
\begin{figure}[t]
\begin{center}
\vspace{-0.1in}
\epsfysize=2.5in
\hspace*{0in}
\epsffile{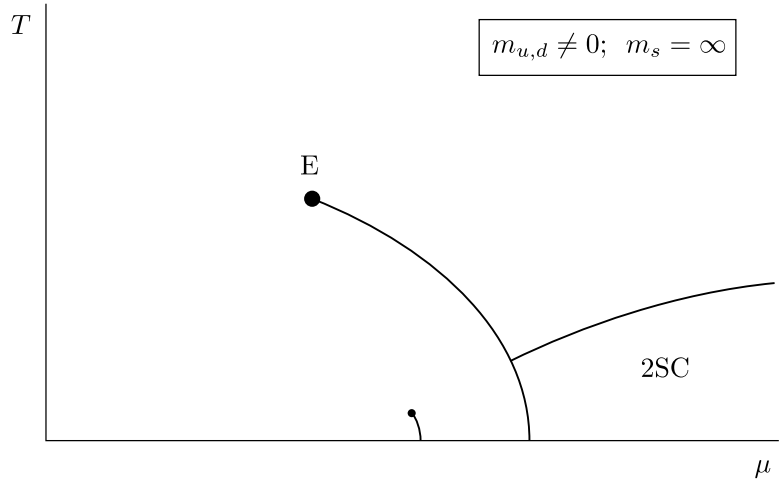}
\end{center} \label{fig2}
\caption{QCD phase diagram for two light quarks. Qualitatively as in Figure~1,
except that the introduction of light quark masses  turns the
second order phase transition into a smooth crossover. 
The tricritical point becomes the critical endpoint $E$, which
can be found in heavy ion collision experiments.}
\vspace{-0.1in}
\end{figure}

Arguments based on a variety of 
models~\cite{NJL,steph,ARW1,RappETC,bergesraj,stephetal}
indicate that the chiral symmetry restoration
transition is first order at 
large $\mu$.  (In Section III, 
we describe the color superconducting (2SC) phase
of cold dense quark matter which occurs at values of $\mu$ 
above this first order transition; the fact that this is a 
transition in which two different condensates compete 
strengthens the argument that this transition
is first order~\cite{bergesraj,CarterDiakonov}.)
This suggests that the
phase diagram features a critical point $E$ at which
the line of first order phase transitions present for 
$\mu>\mu_E$ ends, as shown in Figure~2.\footnote{If
the up and down quarks were massless, $E$ would
be a tricritical point~\cite{lawrie}, at which the first
order transition becomes second order. See Figure~1.}
At $\mu_E$, the phase transition is second order
and is in the Ising universality class~\cite{bergesraj,stephetal}.
Although the
pions remain massive, the correlation length in the $\sigma$ channel
diverges due to universal long wavelength fluctuations
of the order parameter.
This results in characteristic signatures,
analogues of critical opalescence in the sense that they
are unique to collisions which freeze out near the
critical point, which
can be used to discover $E$~\cite{SRS1,SRS2}.   

Returning to the $\mu=0$ axis,
universal arguments~\cite{piswil}, again backed by lattice 
simulation~\cite{latticereview},
tell us that if the strange quark were as light as the
up and down quarks, the transition would be first order,
rather than a smooth crossover.  
This means that if one could dial
the strange quark mass $m_s$, one would find a critical
$m_s^c$ at which the transition as a function of temperature
is second order~\cite{rajwil,rajreview}. 
Figures~2, 3 and 4 are drawn
for a sequence of decreasing strange quark masses. Somewhere
between Figures~3 and 4, $m_s$ is decreased below $m_s^c$ and
the transition on the vertical axis becomes first order.
The value of $m_s^c$ is an open question,
but lattice simulations suggest that it is about half the
physical strange quark mass~\cite{columbia,kanaya}. 
These results are not yet conclusive~\cite{oldkanaya}  but
if they are correct then the phase
diagram in nature is as shown in Figure~3,  and the phase transition
at low $\mu$ 
is a smooth crossover.  
\begin{figure}[t]
\begin{center}
\vspace{-0.1in}
\epsfysize=2.5in
\hspace*{0in}
\epsffile{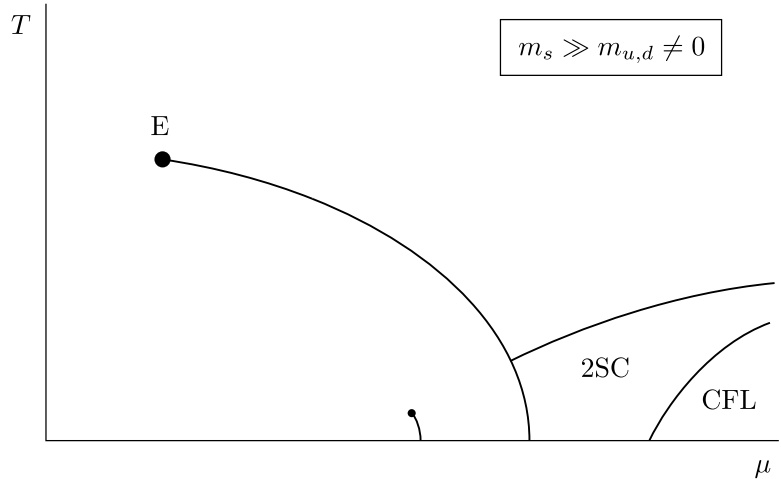}
\end{center} \label{fig3}
\caption{QCD phase diagram for two light quarks and a strange quark
with a mass comparable to that in nature.
The presence of the strange quark shifts
$E$ to the left,  as can be seen by comparing with Figure~2. 
At sufficiently high density, cold quark matter is necessarily 
in the CFL phase in which quarks of all three colors and
all three flavors form Cooper pairs. The diquark condensate in
the CFL phase breaks chiral symmetry, and this
phase has the same symmetries as baryonic matter which is
dense enough that the nucleon and hyperon densities are 
comparable. The phase transition
between the CFL and 2SC phases is first order.}
\vspace{-0.15in}
\end{figure}
\begin{figure}[t]
\begin{center}
\vspace{-0.1in}
\epsfysize=2.5in
\hspace*{0in}
\epsffile{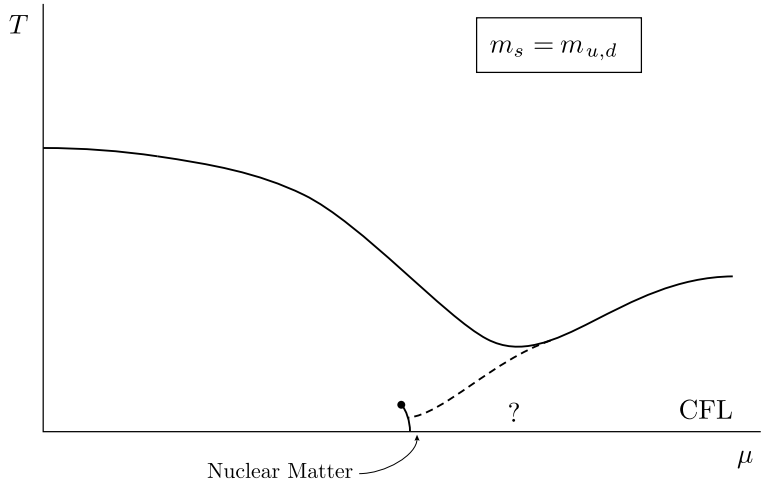}
\end{center} \label{fig4}
\caption{QCD phase diagram for three quarks which
are degenerate in mass and which are either massless or light.
The CFL phase
and the baryonic phase have the same symmetries and 
may be continuously connected. The dashed line 
denotes the critical temperature at which baryon-baryon (or
quark-quark) pairing vanishes; the region below the dashed line 
is superfluid.
Chiral symmetry is 
broken everywhere below the solid line, which is a first order
phase transition. 
The question mark serves to remind
us that although no transition is required in this region,
transition(s) may nevertheless arise as the magnitude of the gap
increases qualitatively in going from the hypernuclear to the
CFL phase.  For quark
masses as in nature, the high density region of the map
may be as shown in Figure~3 or may be closer to that shown here, albeit
with transition(s) in the vicinity of the question mark
associated with the onset
of nonzero hyperon density and the breaking 
of $U(1)_S$ \protect\cite{ABR2+1}.}
\vspace{-0.15in}
\end{figure}

These observations fit together in a simple
and elegant fashion.
If we
could vary  $m_s$, we would find that as $m_s$
is reduced from infinity to $m_s^c$, the critical
point $E$ in the $(T,\mu)$ plane moves toward the $\mu=0$
axis~\cite{SRS1}.  This is shown in Figures~2-4.
In nature, $E$ is at some nonzero $T_E$ and $\mu_E$.
When $m_s$ is reduced to $m_s^c$, between Figure~3 and Figure~4,
$\mu_E$ reaches zero.
Of course, experimentalists cannot vary $m_s$.  They
can, however, vary $\mu$.  AGS collisions with
center of mass energy 
$\sqrt{s}=5$~AGeV create fireballs which freeze out
near $\mu\sim 500-600$~MeV~\cite{PBM}.  
SPS collisions with $\sqrt{s}=17$~AGeV
create fireballs which freeze out near $\mu\sim 200-300$~MeV~\cite{PBM}.
In time, we will also have data from SPS collisions
with $\sqrt{s}=9$~AGeV and from RHIC collisions with
$\sqrt{s}=56$, 130 and 200~AGeV and other energies.\footnote{The 
first data from RHIC collisions at $\sqrt{s}=56$~AGeV and 
$\sqrt{s}=130$~AGeV
have already appeared~\cite{PHOBOS}. 
This bodes well for the analyses to come.} 
By dialing $\sqrt{s}$ and thus $\mu$, experimenters can find
the critical point $E$.  

\section{Discovering the Critical Point}

We hope that the study of heavy ion collisions will, in the
end, lead both to a quantitative study of the 
properties of the quark-gluon plasma
phase at temperatures well above the transition and to a 
quantitative understanding of how to draw the phase transition
region of the phase diagram.  Probing the partonic matter
created early in the collision relies on a suite of signatures
including: the use of $J/\Psi$ mesons, charmed mesons, and perhaps
the $\Upsilon$ as probes; the energy loss of high momentum
partons and consequent effects on the high-$p_T$ hadron spectrum;
and the detection of photons and dileptons over and above those 
emitted in the later hadronic stages of the collision.
I will not review this program here.  Instead, I focus
on signatures of the critical point.
The map of the QCD phase diagram which I have sketched
so far is simple, coherent and consistent with all we know
theoretically; the discovery of the critical point would
provide an experimental foundation for the central
qualitative feature of the landscape.
This discovery would in addition
confirm that in higher energy heavy ion collisions and in
the big bang, the QCD phase transition is a smooth crossover.
Furthermore, the 
discovery of collisions
which create matter that
freezes out near $E$ would imply that 
conditions above the transition existed prior to freezeout,
and would thus make it much easier to interpret the results of other
experiments which study those observables which can probe the 
partonic matter created early in the collision. 

We theorists must clearly do as much as we can
to tell experimentalists {\it where} and {\it how} to find $E$.
The ``where'' question, namely the question of predicting
the value of $\mu_E$ and thus suggesting the $\sqrt{s}$ to
use to find $E$, is much harder for us to answer.   
First, as we stress further in the next Section, 
{\it ab initio} analysis of QCD in its full glory --- i.e.
lattice calculations --- are at present impossible at nonzero $\mu$.
We must therefore rely on models.
Second, an intrinsic feature of the picture we have described 
is that $\mu_E$ is sensitive to the mass of the strange quark,
and therefore particularly hard to predict.
Crude models suggest that $\mu_E$ could be $\sim 600-800$~MeV
in the absence of the strange quark~\cite{bergesraj,stephetal}; 
this in turn suggests that
in nature $\mu_E$ may have of order half this value, and may therefore
be accessible at the SPS if the SPS 
runs with $\sqrt{s}<17$~AGeV.   However, at present theorists cannot
predict the value of $\mu_E$ even to within a factor of two.
The SPS can search a significant fraction of the parameter
space; if it does not find $E$, it will then be up to 
the RHIC experiments to map the $\mu< 200 $~MeV region.

Although we are trying
to be helpful with the ``where'' question, we are not very
good at answering it quantitatively.  This question can only
be answered convincingly by an experimental discovery.
What we theorists {\it can} do reasonably
well is to answer the ``how'' question, thus
enabling experimenters to answer ``where''.  
This is the goal of a recent paper by Stephanov, myself
and Shuryak~\cite{SRS2}.
The signatures we have proposed are based
on the fact that $E$ is a genuine thermodynamic singularity
at which susceptibilities diverge and the order parameter
fluctuates on long wavelengths. The resulting signatures
are {\it nonmonotonic} as a function of $\sqrt{s}$: as
this control parameter is varied, we should see the signatures
strengthen and then weaken again as the critical point is
approached and then passed.   

The critical point $E$ can also be sought by varying control
parameters other than $\sqrt{s}$. 
Ion size, centrality selection
and rapidity selection can all be varied.
The advantage of using $\sqrt{s}$  
is that we already know (by comparing
results from the AGS and SPS)
that dialing it changes 
the freeze out chemical potential $\mu$, which is the goal
in a search for $E$.

The simplest observables we
analyze are the event-by-event fluctuations of the 
mean transverse momentum of the charged particles
in an event, $p_T$, and of the total charged multiplicity in an event, $N$.
We calculate the magnitude of the effects of critical
fluctuations on these and other observables, making
predictions which, we hope, will allow experiments to find $E$.
As a necessary prelude, we analyze the contribution of
noncritical thermodynamic fluctuations. 
We compare the
noncritical fluctuations of an equilibriated resonance gas
to the fluctuations measured by NA49 at $\sqrt{s}=17$~AGeV~\cite{NA49}. 
The observed
fluctuations are as perfect Gaussians as the data statistics
allow, as expected for freeze-out from a system in thermal equilibrium.
The data on multiplicity fluctuations show evidence for
a nonthermodynamic contribution, which is to 
be expected since the extensive quantity $N$ is sensitive
to the initial size of the system and thus to nonthermodynamic
effects like variation in impact parameter. The
contribution of such effects to the fluctuations
have now been estimated~\cite{BH,DS}; the combined thermodynamic
and nonthermodynamic fluctuations are in satisfactory agreement
with the data~\cite{DS}.
The width of the event-by-event distribution\footnote{This width 
can be measured even
if one observes only two pions per event~\cite{bialaskoch};
large acceptance data as from NA49 is required in order
to learn that the distribution is Gaussian, that 
thermodynamic predictions may be valid, and that
the width is therefore the only interesting quantity to measure.} 
of mean $p_T$
is in good agreement with predictions based on noncritical thermodynamic 
fluctuations.
That is, NA49 data are consistent with the hypothesis
that almost all the observed event-by-event fluctuation in 
mean $p_T$, an intensive quantity,
is thermodynamic in origin.
This bodes well for the detectability of systematic
changes in thermodynamic fluctuations near $E$.

One 
analysis described in detail in Ref.~\cite{SRS2} is based on the ratio
of the width of the true event-by-event distribution of the mean $p_T$
to the width of the distribution in a sample of mixed events. This
ratio was called $\sqrt{F}$. NA49 has measured $\sqrt{F}=1.002\pm
0.002$~\cite{NA49,SRS2}, which is consistent with expectations for
noncritical thermodynamic fluctuations.\footnote{In an infinite system
made of classical particles which is in thermal equilibrium,
$\sqrt{F}=1$.  Bose effects increase $\sqrt{F}$ by $1-2\%$
\cite{Mrow,SRS2}; an anticorrelation introduced by energy conservation
in a finite system --- when one mode fluctuates up it is more likely
for other modes to fluctuate down --- decreases $\sqrt{F}$ by $1-2\%$
\cite{SRS2}; two-track resolution also decreases $\sqrt{F}$ by $1-2\%$
\cite{NA49}. The contributions due to correlations introduced by
resonance decays and due to fluctuations in the flow velocity are each
much smaller than $1\%$~\cite{SRS2}.}  Critical fluctuations
of the $\sigma$ field, i.e. the characteristic long wavelength
fluctuations of the order parameter near $E$, influence pion momenta
via the (large) $\sigma\pi\pi$ coupling and increase $\sqrt{F}$
\cite{SRS2}.  The effect is proportional to $\xi_{\rm freezeout}^2$,
where $\xi_{\rm freezeout}$ is the $\sigma$-field correlation length
of the long-wavelength fluctuations at freezeout~\cite{SRS2}.  If
$\xi_{\rm freezeout}\sim 3$ fm (a reasonable estimate, as we
describe below)
the ratio $\sqrt{F}$ increases by
$\sim 3-5\%$, ten to twenty times the statistical error in the present
measurement~\cite{SRS2}.  This observable is valuable because data on
it has been analyzed and presented by NA49, and it can therefore be
used to learn that Pb+Pb collisions at 158~AGeV do {\it not} freeze
out near~$E$. The $3-5\%$ nonmonotonic
variation in $\sqrt{F}$ as a function
of $\sqrt{s}$  which we predict is easily detectable but is
not so large as to make one confident of using this alone
as a signature of $E$.

Once $E$ is located, however, other observables which 
are more sensitive to critical effects will be more useful.
For example, a $\sqrt{F_{\rm soft}}$,
defined using only the softest $10\%$ of the pions in each event, 
will be much more sensitive to the critical long wavelength 
fluctuations.  The higher $p_T$ pions are less affected
by the $\sigma$ fluctuations~\cite{SRS2}, 
and these relatively unaffected pions
dominate the mean $p_T$ of all the pions in the
event.  This is why the increase in $\sqrt{F}$ near the critical point 
will be much less than that of $\sqrt{F_{\rm soft}}$. 
Depending on the details of the
cuts used to define it,  $\sqrt{F_{\rm soft}}$ should be enhanced by many tens
of percent in collisions passing near $E$.
Ref.~\cite{SRS2} suggests other such observables,
and more can surely be found.

The multiplicity of soft pions is an 
example of an observable which may
be used to detect the critical fluctuations 
without an event-by-event analysis.
The post-freezeout decay of sigma mesons, which are copious
and light at freezeout near $E$ and which
decay subsequently when their mass increases above
twice the pion mass, should result in a population of pions 
with $p_T\sim m_\pi/2$ which appears only for freezeout
near the critical point~\cite{SRS2}.  
If $\xi_{\rm freezeout}> 1/m_\pi$, this population
of unusually low momentum pions will be comparable in
number to that of the ``direct'' pions (i.e. those which
were pions at freezeout) and will result in a large
signature.  This signature is therefore certainly
large for $\xi_{\rm freezeout}\sim 3$~fm and would
not increase much further if $\xi_{\rm freezeout}$ were larger still.

The variety of observables
which should {\it all} vary nonmonotonically with $\sqrt{s}$
(and should all peak at the same $\sqrt{s}$)
is sufficiently great that if it were to turn out that 
$\mu_E<200$~MeV, making $E$ inaccessible to the SPS, all four
RHIC experiments could play a role in the study of the critical
point.

The purpose of Ref.~\cite{Berdnikov} 
is to estimate how large 
$\xi_{\rm freezeout}$ can become, thus making the predictions
of Ref.~\cite{SRS2} for the magnitude of various signatures
more quantitative.  
The nonequilibrium dynamics analyzed in Ref.~\cite{Berdnikov}
is guaranteed
to occur in a heavy ion collision which passes near $E$,
even if local thermal equilibrium is achieved 
at a higher temperature during the earlier evolution
of the plasma created in the collision.  
If this plasma were to cool arbitrarily slowly, $\xi$ would
diverge at $T_E$.  However, it would take an infinite
time for $\xi$ to grow infinitely large.  Indeed, near
a critical point, the longer the correlation length, the
longer the equilibration time, and the slower the 
correlation length can grow. This critical slowing
down means that the
correlation length cannot grow sufficiently fast for the 
system to stay in equilibrium.  
We use the theory of dynamical critical phenomena to 
describe the effects of critical slowing down
of the long wavelength dynamics near $E$
on the time development of the correlation length.
The correlation length does not have time
to grow as large as it would in equilibrium:  
we find 
$\xi_{\rm freezeout}\sim 2/T_E \sim 3$ fm for trajectories
passing near $E$. 
Although critical slowing down hinders the growth of $\xi$, it 
also slows the decrease of $\xi$  as the system continues to cool 
below the critical point.  As 
a result, $\xi$ does not
decrease significantly between the phase transition and 
freezeout.  

Our results depend on the universal 
function describing the equilibrium
behavior of $\xi$ near the Ising critical point $E$, 
on the universal dynamical exponent
$z$ describing critical slowing down (perturbations away from
equilibrium relax toward equilibrium on a timescale which
scales with $\xi$ like $A\xi^z$~\cite{HoHa}), on the nonuniversal
constant $A$, the nonuniversal constants  
which relate  $(T-T_E)$ and $(\mu-\mu_E)$ to dimensionless
Ising model variables, on $T_E$
which we take to be $\sim 140$~MeV,
and finally on the cooling rate $|dT/dt|$ which
we estimate to be 4 MeV/fm~\cite{Bravina,Berdnikov}.

Our estimate that $\xi$ does not
grow larger than $2/T_E$ is robust in three senses.  First, it depends
very little on the angle with which the trajectory passes
through $E$. Second, 
it turns out to depend on only one combination of all the
nonuniversal quantities which play a role. We call this
parameter $a$; it is proportional to $|dT/dt|^{-1}$. 
Third, our results do not depend sensitively on $a$. 
We show that the maximum value of $\xi$ scales like $a^{\frac{\nu/\beta\delta}
{1+z\nu/\beta\delta}}\approx a^{0.215}$ 
\cite{Berdnikov}.\footnote{A scaling law
of this form (of course with different numerical values for the exponents) 
relating the maximum correlation length which is
reached to the cooling rate 
was first discovered in the theory of defect formation at a 
second order phase transition~\cite{Zurek}. It has been tested
in this context in
numerical simulations~\cite{Antunes} and, furthermore, is supported by 
data from experiments on
liquid crystals~\cite{LiquidCrystal} and superfluid $^3$He~\cite{Helium}.}
Thus, for example,
$|dT/dt|$ would have to be a factor of 25 smaller than 
we estimate in order
for $\xi$ to grow to $4/T_E$ instead of $2/T_E$.
Although our results are robust in this sense,
they cannot be treated as precise because 
our assumption that the dynamics of $\xi$ in QCD 
is described by the universal classical dynamics
of the three-dimensional Ising model only becomes precise if $\xi\gg 1/T_E$,
while our central result is that $\xi$ does not grow beyond $\sim 2/T_E$.
A $3+1$-dimensional quantum field theoretical treatment of 
the interplay between cooling and 
the dynamics of critical slowing down is not
yet available, but promising first steps in this direction 
can be found in Ref.~\cite{Boyan}.

A result which is of great importance in the planning
of experimental searches 
is that one need not hit $E$ precisely in order
to find it.  Our analysis demonstrates that if one were to
do a scan with collisions at many finely spaced values of
the energy and thus $\mu$, one would see signatures of $E$
with approximately the same magnitude over a broad range
of $\mu$.  The magnitude of the 
signatures will not be narrowly peaked as $\mu$ is varied.
As long as one gets close enough to $E$ that the equilibrium
correlation length is $(2-3)/T_E$, the actual correlation
length $\xi$ will grow to $\sim 2/T_E$.  There is no advantage
to getting closer to $E$, because critical slowing down
prevents $\xi$ from getting much larger even if $\xi_{\rm eq}$ does.
Data at many finely spaced values of $\mu$ is
{\it not} called 
for.

As described above,
knowing that we are
looking for  $\xi_{\rm freezeout} \sim 3$~fm allows us \cite{Berdnikov}
to make quantitative estimates of the magnitude of the signatures
of $E$ described in detail in Ref. \cite{SRS2}.
Together, the excess
multiplicity at low momentum (due to post-freezeout sigma decays) and
the excess event-by-event fluctuation of the momenta of
the low momentum pions (due to their coupling to the order parameter
which is fluctuating with correlation length $\xi_{\rm freezeout}$)
should allow a convincing detection of the critical point $E$.
Both should behave nonmonotonically as the collision energy,
and hence $\mu$, are varied.  Both should peak for those
heavy ion collisions which freeze out near $E$, with
$\xi_{\rm freezeout}\sim 3$ fm.

We have learned much from the beautiful gaussian event-by-event
fluctuations observed by NA49.  The magnitude of these fluctuations
are consistent with the hypothesis that the hadronic system
at freezeout is in approximate thermal equilibrium. These and
other data show none of the non-gaussian features that would 
signal that the system had
been driven far from equilibrium either by a rapid
traversal of the transition region or by the bubbling
that would occur near a strong first order
phase transition.  There is also no sign of
the enhanced, but still gaussian, fluctuations which would signal
freezeout near the critical point $E$.  Combining these 
observations with the observation of tantalizing indications
that the matter created in SPS collisions is not well described
at early times by hadronic models~\cite{HeinzJacob} suggests
that collisions at the SPS may be exploring the crossover region
to the left of the critical point $E$, in which
the matter is not well-described as a hadron gas but
is also not well-described as a quark-gluon plasma.  This speculation
could be confirmed in two ways.   First, if the SPS is probing
the crossover region then the coming experiments
at RHIC may discover direct signatures of an early partonic phase, 
which are well-described by theoretical calculations beginning from 
an equilibrated quark-gluon plasma.  
Second, 
if $\sqrt{s}=17$~AGeV collisions
are probing the
crossover region not far to the left of the critical point $E$, 
then SPS data taken at lower energies
would result in the discovery of $E$. 
If, instead, RHIC were to discover
$E$ with $\mu_E<200$~MeV, that would indicate that the SPS 
experiments have probed the weakly first order region 
just to the right of $E$. Regardless, 
discovering $E$ would take all the speculation
out of mapping this part of the QCD phase diagram.

\section{Color Superconductivity and Color-Flavor Locking}

I turn now
to recent developments in our understanding
of the low temperature, high density
regions of the QCD phase diagram.  
First, a notational confession.  It is conventional 
in the literature on cold dense quark matter to define $\mu$
as the {\it quark} number chemical potential, $1/3$ the baryon
number chemical potential used in Sections I and II.  We make
this change from here on.  For example, neutron star cores
likely have $\mu\sim 400-500$~MeV, corresponding
to baryon number chemical potentials $\sim 1.2-1.5$~GeV in
Figures~1-4.  

The relevant degrees
of freedom in cold dense quark matter
are those which involve quarks with momenta
near the Fermi surface.  At high density, when the 
Fermi momentum is large, the QCD gauge coupling $g(\mu)$ is small.
However, because of the infinite degeneracy among  
pairs of quarks with equal and opposite momenta
at the Fermi surface, 
even an arbitrarily weak attraction between quarks renders
the Fermi surface unstable to the formation of a condensate
of quark Cooper pairs.  
Creating a pair costs no free energy 
at the Fermi surface and the attractive interaction
results in a free energy benefit.
Pairs of quarks cannot be color
singlets, and in QCD with two flavors of massless quarks
the Cooper pairs form in the (attractive) color ${\bf \bar 3}$ 
channel~\cite{Barrois,BailinLove,ARW1,RappETC}.
The resulting 
condensate creates a gap $\Delta$ at the Fermi surfaces of
quarks with two
out of the three colors and breaks $SU(3)_{\rm color}$ to an 
$SU(2)_{\rm color}$ subgroup, giving mass to five of
the gluons by the Anderson-Higgs mechanism.
In QCD with two flavors, the Cooper pairs are $ud-du$
flavor singlets and the global flavor symmetry 
$SU(2)_L\times SU(2)_R$ is intact. There
is also an unbroken global symmetry which plays the
role of $U(1)_B$. Thus, no global symmetries are broken
in this 2SC phase.  There must therefore be a phase
transition between the 2SC and hadronic
phases on the horizontal axis in Figure~1, at which
chiral symmetry is restored.  This phase transition
is first order~\cite{ARW1,bergesraj,PisarskiRischke1OPT,CarterDiakonov}
since it involves
a competition between chiral condensation and
diquark condensation~\cite{bergesraj,CarterDiakonov}.
There need be no transition between the 2SC and quark-gluon
plasma phases in Figure~1 because neither phase breaks any global
symmetries.
However, this transition, which
is second order in mean field theory, is likely first
order in QCD due to gauge field fluctuations~\cite{bergesraj}, 
at least at high enough 
density~\cite{PisarskiPhaseDiagram}.

In QCD with three flavors of massless quarks, the Cooper
pairs {\it cannot} be flavor singlets, and both color and flavor
symmetries are necessarily broken. The symmetries of
the phase which results have been analyzed 
in~\cite{CFL,SW1}.  The attractive channel favored
by one-gluon exchange exhibits ``color-flavor locking.''
A condensate 
of the form 
\begin{equation}
\label{CFLform}
\langle \psi_L^{\alpha a}\psi_L^{\beta b}\rangle 
\propto \Delta \epsilon^{\alpha\beta A}\epsilon^{abA} 
\end{equation}
involving left-handed quarks alone, 
with $\alpha$, $\beta$ color indices and $a$, $b$ flavor indices,
locks $SU(3)_L$ flavor rotations to $SU(3)_{\rm color}$:
the condensate is not symmetric under either alone, but is
symmetric under the 
simultaneous $SU(3)_{L+{\rm color}}$ rotations.\footnote{It turns
out~\cite{CFL} that condensation in the color ${\bf \bar 3}$ channel
induces a condensate in the color ${\bf 6}$ channel 
because this breaks no further symmetries~\cite{ABR2+1}.
The resulting condensates can be written in terms 
of $\kappa_1$ and $\kappa_2$ where
$\langle \psi^{\alpha a}_L \psi^{\beta b}_L \rangle \sim \kappa_1 
\delta^{\alpha a}
\delta^{\beta b} + \kappa_2 \delta^{\alpha b} \delta^{\beta a}$. Here,
the Kronecker $\delta$'s
lock color and flavor rotations. The pure color ${\bf \bar 3}$ 
condensate (\ref{CFLform}) has $\kappa_2=-\kappa_1$.}
A condensate involving right-handed quarks alone 
locks $SU(3)_R$ flavor rotations to $SU(3)_{\rm color}$.
Because color is vectorial, the combined effect
of the $LL$ and $RR$ condensates is to lock $SU(3)_L$
to $SU(3)_R$, breaking chiral symmetry.\footnote{Once 
chiral symmetry is broken by color-flavor locking, 
there is no symmetry argument precluding the existence
of an ordinary chiral condensate. Indeed,
instanton effects do induce a nonzero $\langle \bar q q \rangle$~\cite{CFL},
but this is a small effect~\cite{RappETC2}.}
Thus, in quark matter with three massless quarks,
the $SU(3)_{\rm color}\times SU(3)_L \times SU(3)_R \times U(1)_B$
symmetry is broken down to the global diagonal $SU(3)_{{\rm color}+L+R}$
group.  A gauged $U(1)$ subgroup 
of the original symmetry group --- a linear combination of one
color generator and electromagnetism, which lives within
$SU(3)_L \times SU(3)_R$ --- also remains unbroken.  
All nine quarks have a gap. All eight gluons get a mass.
There are nine massless Nambu-Goldstone bosons.
All the quarks, all the massive vector bosons, and
all the Nambu-Goldstone bosons have integer charges
under the unbroken gauged $U(1)$ symmetry, which therefore plays
the role of electromagnetism.
The CFL phase therefore has the same symmetries
as baryonic matter with a condensate of
Cooper pairs of baryons~\cite{SW1}.  Furthermore,
many non-universal features of these two phases 
correspond~\cite{SW1}.
This raises the possibility that quark matter and baryonic
matter may be continuously connected~\cite{SW1}, 
as shown in Figure~4.  

The physics of 
the CFL phase has been the focus of much recent 
work~\cite{CFL,SW1,ABR2+1,SW2,RappETC2,Zahed,effectiveCFL,gapless,SchaeferPatterns,SonStephMesons,RWZ,HongLeeMin,ManuelTytgat,RSWZ,Zarembo,BBSMeson,RischkeMeissner,HongEMMass,SchaeferKCond,Nowak}.
Nature chooses two light quarks and one middle-weight
strange quark, rather than three
degenerate quarks as in Figure~4. 
A nonzero $m_s$ weakens those condensates which
involve pairing between light and strange quarks.
The CFL phase requires
nonzero $\langle us \rangle$ and $\langle ds \rangle$
condensates; because these condensates
pair quarks with differing Fermi momenta 
they can only exist if they are 
larger than of order $m_s^2/2\mu$, the
difference between the $u$ and $s$ Fermi momenta in
the absence of pairing.
If one imagines increasing $m_s$ at fixed $\mu$, one finds a first order
unlocking transition~\cite{ABR2+1,SW2}: for larger
$m_s$ only $u$ and $d$
quarks pair and the 2SC phase is obtained.  
Conversely, as $m_s$
is reduced in going from Figure~2 to 3 to 4, the 
region occupied by the CFL phase expands to encompass 
regions with smaller
and smaller $\mu$~\cite{ABR2+1,SW2}.  For  
any $m_s\neq \infty$,
the CFL phase is the ground state at arbitrarily
high density~\cite{ABR2+1}.  For larger values of $m_s$,
there is a 2SC interlude on the horizontal axis,
in which chiral symmetry is restored, before
the CFL phase breaks it again at high densities.
For smaller values of $m_s$, the possibility of
quark-hadron continuity~\cite{SW1} as shown in Figure~4 arises.
It should be noted that even when the strange and light quarks 
are not degenerate,
the CFL phase may be continuous
with a baryonic phase in which the densities of
all the nucleons and hyperons are comparable; there
are, however, phase transitions between this 
hypernuclear phase and ordinary nuclear matter~\cite{ABR2+1}.

The Nambu-Goldstone bosons in the CFL phase
are Fermi surface excitations in which the orientation
of the left-handed and right-handed diquark condensates oscillate
out of phase in flavor space.
The effective field theory describing
these oscillations has been 
constructed~\cite{effectiveCFL,SonStephMesons,Zarembo}.  
Because the full theory is weakly coupled at asymptotically
high densities, in this regime all coefficients in the effective theory
describing the long wavelength meson physics are calculable
from first principles.  The decay 
constants $f_{\pi,K,\eta,\eta'}$~\cite{SonStephMesons}
and the meson masses 
$m_{\pi,K,\eta,\eta'}$~\cite{SonStephMesons,RWZ,HongLeeMin,ManuelTytgat,RSWZ,BBSMeson} are all now known.  
The meson masses depend on quark masses like $m^2\sim m_q^2$
in the CFL phase (neglecting the small chiral condensate)~\cite{CFL}, and
their masses are inverted in the sense that the kaon is lighter
than the pion~\cite{SonStephMesons}.  The charged kaon mass
$m_{K^\pm}^2\sim m_d(m_u+m_s)\Delta/\mu$ is so light that 
it is likely less than the electron chemical
potential, meaning that the CFL phase 
likely features a kaon condensate~\cite{SchaeferKCond}.
The dispersion 
relations describing the fermionic quasiparticle excitations in
the CFL phase, which have the quantum numbers of an octet and a singlet
of baryons,  
have also received attention~\cite{ABR2+1,gapless}.
So have the properties of the massive vector meson octet ---
the gluons which receive a mass via the Meissner-Anderson-Higgs 
mechanism~\cite{SonStephMesons,TwoFlavorMeissner,RischkeMeissner}.
We now have a description of the properties
of the CFL phase and its excitations, in which much
is known quantitatively if the value of the gap $\Delta$
is known.  We describe estimates of $\Delta$ below.

It is interesting that both the 2SC and CFL phases satisfy
anomaly matching constraints, even though
it is not yet completely clear whether this
must be the case when Lorentz invariance
is broken by a nonzero density~\cite{AnomalyMatching}.
It is not yet clear how high density 
QCD with larger numbers of flavors~\cite{SchaeferPatterns} satisfies
anomaly matching constraints.
Also, anomaly matching in the 2SC phase requires that the
up and down quarks of the third color remain ungapped; 
this requirement must, therefore,
be modified once these quarks pair to form a 
$J=1$ condensate, breaking rotational invariance~\cite{ARW1}. 

Much effort has gone into estimating
the magnitude of the gaps in the 2SC and CFL 
phases~\cite{BailinLove,ARW1,RappETC,CFL,ABR2+1,SW2,bergesraj,CarterDiakonov,RappETC2,Hsu1,SW0,AKS,Vanderheyden,Son,PisarskiRischke,Hong,HMSW,SW3,rockefeller,Hsu2,ShovWij,EHHS,BBS,RajagopalShuster,Manuel}.
It would be ideal if this task were within the scope of
lattice gauge theory as is, for example, the calculation
of the critical temperature on the vertical axis of the phase diagram.
Unfortunately, lattice methods relying
on importance sampling have to this
point been rendered exponentially 
impractical at nonzero baryon density by the 
complex action at nonzero $\mu$. There are more 
sophisticated algorithms which
have allowed theories which are simpler than QCD but which
have as severe a fermion sign problem as that in QCD at nonzero 
chemical potential to be simulated~\cite{MeronCluster}.
This bodes well for the future.\footnote{Note that 
quark pairing can be studied on the lattice in some models
with four-fermion interactions and in two-color QCD~\cite{HandsMorrison}. 
The $N_c=2$ case has also
been studied analytically in Refs.~\cite{RappETC,analytic2color}; 
pairing in this
theory is simpler to analyze because 
quark Cooper pairs are color singlets. 
The $N_c\rightarrow \infty$
limit of QCD is often one in which hard problems become
tractable. However, the ground state of $N_c=\infty$ QCD
is a chiral density wave, not a color superconductor~\cite{DGR}.
At asymptotically high densities 
color superconductivity persists up
to $N_c$'s of order thousands~\cite{ShusterSon,PRWZ} before being
supplanted by the phase described in Ref.~\cite{DGR}.  At any finite
$N_c$, color superconductivity occurs at 
arbitrarily weak coupling whereas
the chiral density wave does not.
For $N_c=3$, color superconductivity is 
still favored over the chiral density wave (although not by much)
even if the interaction 
is so strong that the color superconductivity gap is 
$\sim \mu/2$~\cite{RappCrystal}.
}
Given the present absence of suitable lattice methods, the
magnitude of the gaps in quark matter at large but accessible
density has been estimated using two broad strategies.
The first class of estimates are done within the
context of models whose
parameters are chosen to give reasonable vacuum
physics.  Examples include analyses in which the
interaction between quarks is replaced simply by four-fermion
interactions with the quantum numbers of
the instanton interaction~\cite{ARW1,RappETC,bergesraj}
or of one-gluon exchange~\cite{CFL,ABR2+1},
random matrix models~\cite{Vanderheyden}, and more sophisticated
analyses done using
the instanton liquid model~\cite{CarterDiakonov,RappETC2,RappCrystal}.
Renormalization group methods have also been used to
explore the space of all possible 
effective four-fermion interactions~\cite{Hsu1,SW0}.
These methods yield results which are in qualitative 
agreement: the favored condensates are as described
above; the gaps range between several tens of MeV up to as large as about 
$100$~MeV;
the associated critical temperatures (above which the 
diquark condensates vanish)
can be as large as about $T_c\sim 50$~MeV.
This agreement between different models reflects the fact
that what matters most is simply the strength of the attraction
between quarks in the color ${\bf \bar 3}$ channel, and by
fixing the parameters of the model interaction to fit, say,
the magnitude of the  vacuum chiral condensate, one ends up
with attractions of similar strengths in different models.

The second strategy for estimating gaps and critical
temperatures is to use
$\mu=\infty$ physics as a guide.
At asymptotically large $\mu$, models with short-range interactions
are bound to fail because the dominant interaction is due
to the long-range magnetic interaction coming from single-gluon
exchange~\cite{PisarskiRischke1OPT,Son}.  The collinear infrared
divergence in small angle scattering via one-gluon exchange
(which is regulated by dynamical screening~\cite{Son})
results in a gap which is parametrically larger at $\mu\rightarrow\infty$
than it would be for any point-like four-fermion interaction.
At $\mu\rightarrow\infty$, where $g(\mu)\rightarrow 0$,
the gap takes the 
form~\cite{Son}
\begin{equation}
\Delta \sim b \mu \,g(\mu)^{-5} \exp[-3\pi^2/\sqrt{2}g(\mu)]\ ,
\label{eq:SonResult}
\end{equation}
whereas for
a point-like interaction with four-fermion coupling $g^2$
the gap goes like $\exp(-1/g^2)$.
Son's result (\ref{eq:SonResult}) has now been confirmed using a variety of 
methods~\cite{SW3,PisarskiRischke,Hong,HMSW,rockefeller,Hsu2,BBS}.
The ${\cal O}(g^0)$ contribution to the prefactor $b$ in 
(\ref{eq:SonResult}) is not yet fully understood.  It is
estimated to 
be $b \sim 512 \pi^4$ in the 2SC phase and 
$b\sim 512 \pi^4 2^{-1/3} (2/3)^{5/2}$ in the CFL
phase~\cite{SW3,PisarskiRischke,HMSW,rockefeller,Hsu2,ShovWij,SchaeferPatterns}.
However, modifications to the quasiparticle dispersion
relations in the normal (nonsuperconducting; high temperature)
phase~\cite{rockefeller}
and quasiparticle damping effects in the superconducting
phase~\cite{Manuel} both tend to reduce $b$. Also, the value
of $b$ is affected by the choice of the scale at which $g$ is
evaluated in (\ref{eq:SonResult}).  The results of Beane {\it et al.}
demonstrate that $g$ should be evaluated at a $\mu$-dependent scale which is
much lower than $\mu$~\cite{BBS}.  If, by convention,
one instead takes $g$ as $g(\mu)$, then $b$ is significantly enhanced.
Finally, examination of the gauge-dependent (and $g$-dependent)
contributions to $b$ in calculations based on
the one-loop Schwinger-Dyson equation (e.g. 
those of Ref.~\cite{SW3,PisarskiRischke,HMSW,Hsu2})
reveals that they only begin to decrease 
for $g< 0.8$~\cite{RajagopalShuster}.
This means that effects which have to date been
neglected in all calculations (e.g. vertex corrections)
are small corrections to $b$ only for $\mu\gg 10^8$~MeV.  

The phase of $N_c=3$ QCD with nonzero 
isospin density ($\mu_I\neq 0$) and zero baryon density ($\mu=0$) {\it can}
be simulated on the lattice~\cite{SonStephIsospin}.  
Although not physically realizable, it is very interesting
to consider because phenomena arise which are similar
to those occurring at large $\mu$ and, in this context,
these phenomena can be analyzed on the lattice.
In this setting, therefore, lattice simulations 
can be used to
test calculational methods which have also been applied at large $\mu$,
where lattice simulation is unavailable.
Large $\mu_I$ physics features large Fermi surfaces for down quarks
and anti-up quarks, Cooper pairing of down
and anti-up quarks,  and a gap
whose $g$-dependence is as in (\ref{eq:SonResult}), albeit
with a different 
coefficient of $1/g$ in the exponent~\cite{SonStephIsospin}.  
This condensate has the same quantum numbers as the pion
condensate expected at much lower $\mu_I$, which means
that a hypothesis of continuity between hadronic --- in this
case pionic --- and quark matter as a function of $\mu_I$ 
can be tested on the lattice~\cite{SonStephIsospin}.
We henceforth return to the physically
realizable setting in which differences between chemical potentials 
for different species of quarks (e.g. $\mu_I$) are small compared
to $\mu$.

At large enough $\mu$, the differences between $u$, $d$ and 
$s$ Fermi momenta decrease, while the
result (\ref{eq:SonResult}) demonstrates  
that the magnitude of the condensates 
{\it increases} slowly as $\mu\rightarrow\infty$. (As
$\mu\rightarrow\infty$, 
the running coupling $g(\mu)\rightarrow 0$ logarithmically and
the exponential factor in (\ref{eq:SonResult}) goes to zero, but
not sufficiently fast to overcome the growth of $\mu$.)  
This means
that the CFL phase is favored over the 2SC phase 
for $\mu\rightarrow\infty$ for any $m_s\neq \infty$~\cite{ABR2+1}.
If we take the asymptotic estimates for the prefactor,
quantitatively valid for $\mu\gg 10^8$~MeV~\cite{RajagopalShuster},
and apply them at accessible densities,
say $\mu\sim 500$~MeV, it predicts gaps as large as 
about $100$~MeV and
critical temperatures as large as about $50$~MeV~\cite{SW3}. 
Even though the asymptotic regime where $\Delta$ can
be calculated from first principles with confidence is
not accessed in nature, it is of great theoretical 
interest.
The weak-coupling calculation of the gap in the CFL phase 
is the
first step toward the weak-coupling calculation of other
properties of this phase, in which chiral symmetry is broken
and the spectrum of excitations is as in a confined phase.
As we have described  above, for example, the masses and decay constants
of the pseudoscalar mesons can be calculated from first
principles once $\Delta$ is known.

It is satisfying that two very different approaches,
one using zero density phenomenology to normalize models, the
other using weak-coupling methods valid at asymptotically
high density, yield predictions for the
gaps and critical temperatures at accessible
densities
which are in good agreement.  Neither can be trusted quantitatively
for quark number chemical potentials $\mu\sim 400-500$~MeV,
as appropriate for the quark matter which may occur in
compact stars.  Still, both methods agree that the
gaps at the Fermi surface are of order tens to 100~MeV, with
critical temperatures about half as large.

$T_c\sim 50$~MeV is much larger relative to the
Fermi momentum (say $\mu\sim 400-500$~MeV) than in 
low temperature superconductivity in metals.
This reflects the fact that color superconductivity
is induced by an attraction due to the primary,
strong, interaction in the theory, rather
than having to rely on much weaker secondary interactions,
as in phonon mediated superconductivity in metals.
Quark matter is a high-$T_c$ superconductor by any reasonable
definition. It is unfortunate
that its $T_c$ is nevertheless low enough that
it is unlikely the phenomenon can be realized in heavy ion
collisions.   

\section{Color Superconductivity in Compact Stars}

Our current understanding of the color superconducting
state of quark matter leads us to believe that it
may occur naturally in compact stars. 
The critical temperature $T_c$ below which quark matter 
is a color superconductor is high enough that
any quark matter which occurs within
neutron stars that are more than a few seconds old
is in a color superconducting state.
In the absence of lattice simulations, present theoretical
methods are not accurate enough to determine whether 
neutron star cores are made of hadronic matter or quark
matter.  They also cannot determine whether any quark
matter which arises will be in the CFL or 2SC phase: 
the difference between the $u$, $d$ and $s$ Fermi momenta
will be a few tens of MeV which is comparable to estimates
of the gap $\Delta$; the CFL phase occurs when $\Delta$ is
large compared to all differences between Fermi momenta.
Just as the higher temperature regions of the QCD
phase diagram are being mapped out in heavy ion collisions,
we need to learn how to use neutron star phenomena to 
determine whether they feature cores made of 2SC quark matter,
CFL quark matter or hadronic matter, thus teaching us
about the high density region of the QCD phase diagram.
It is therefore important to look for astrophysical consequences of
color superconductivity.

{\bf Equation of State:} Much of the work on the consequences 
of quark matter within a compact star has focussed on
the effects of quark matter on the equation of state,
and hence on the radius of the star.  As a Fermi surface
phenomenon, color superconductivity has little effect on
the equation of state: the pressure is an integral over
the whole Fermi volume.  Color superconductivity 
modifies the equation of state at the $\sim (\Delta/\mu)^2$
level, typically by a few percent~\cite{ARW1}.  Such small effects
can be neglected in present calculations, and for
this reason I will not attempt to survey
the many ways in which observations of neutron stars
are being used to constrain the equation of state~\cite{Henning}.

I will describe one current idea, however.
As a neutron star in a low mass X-ray binary (LMXB)
is spun up by accretion from its companion, it becomes
more oblate and its central density decreases. If it contains
a quark matter core, the volume fraction occupied by this
core decreases, the star expands, and its moment of inertia
increases.  This raises the possibility~\cite{GlendenningWeberSpinup}
of a period during the spin-up history of an LMXB when
the neutron star is gaining angular momentum via accretion,
but is gaining sufficient moment of inertia that its angular
frequency is hardly increasing.  In their modelling of this effect,
Glendenning and Weber~\cite{GlendenningWeberSpinup} discover 
that LMXB's should spend a significant fraction
of their history with a frequency of around 200~Hz,
while their quark cores are being spun out of existence,
before eventually spinning up to higher frequencies.  
This may explain the observation that 
LMXB frequencies are clustered around 250-350~Hz~\cite{vanderKlis},
which is otherwise puzzling in that it is thought that LMXB's provide
the link between canonical pulsars and millisecond pulsars,
which have frequencies as large as 600~Hz~\cite{ChakrabartyMorgan}.
It will be interesting to see how robust the result of 
Ref.~\cite{GlendenningWeberSpinup} is to changes in model
assumptions and also how 
its predictions fare when compared to 
those of other 
explanations which posit upper bounds on LMXB 
frequencies~\cite{Bildsten2},
rather than a most probable frequency range with no 
associated upper bound~\cite{GlendenningWeberSpinup}. 
We note here that because Glendenning
and Weber's  effect depends only 
on the equation of state and not on other
properties of quark matter, the fact that the quark
matter must in fact be a color superconductor
will not affect the results in any significant way.
If Glendenning and Weber's explanation for the observed clustering
of LMXB frequencies proves robust, it would imply
that pulsars with lower rotational frequencies feature quark matter
cores.  

{\bf Cooling by Neutrino Emission}: 
We turn now to neutron star phenomena which {\it are} affected
by Fermi surface physics.  For the first $10^{5-6}$ years of its
life, the cooling of a neutron star is governed by the balance
between heat capacity and the loss of heat by neutrino emission.
How are these quantities affected by the presence of a
quark matter core? This has been addressed recently in 
Refs.~\cite{Blaschke,Page}, following earlier work in Ref.~\cite{Schaab}.
Both the specific heat $C_V$  and the neutrino emission rate 
$L_\nu$ are dominated
by physics within $T$ of the Fermi surface.  If, as 
in the CFL phase,  all quarks have a gap $\Delta\gg T$ then
the contribution of quark quasiparticles to $C_V$ and $L_\nu$ 
is suppressed by $\sim \exp(-\Delta/T)$.  There may be other
contributions to $L_\nu$~\cite{Blaschke}, but these are also
very small.  The specific heat is  
dominated by that of the electrons, although it 
may also receive a small contribution from the CFL phase Goldstone
bosons.  Although further work is required, it is already
clear that both $C_V$ and $L_\nu$ are much smaller than in
the nuclear matter outside the quark matter core. This
means that the total heat capacity and
the total neutrino emission rate (and hence
the cooling rate) of a neutron star with a CFL core will 
be determined completely by the nuclear matter outside 
the core.  The quark matter core is ``inert'':
with its small heat capacity and emission rate it
has little influence on the temperature of the star as a whole.
As the rest of the star emits neutrinos and cools, the core
cools by conduction, because the electrons keep it in good thermal
contact with the rest of the star.   These qualitative expectations
are nicely borne out in the calculations presented
by Page et al.~\cite{Page}.

The analysis of the cooling history of a neutron star with 
a quark matter core in the 2SC phase is more complicated.
The red and green up and down quarks pair with a gap
many orders of magnitude larger than the temperature, which is
of order 10 keV, and
are therefore inert as described above.  
Any strange quarks present will form a
$\langle ss \rangle$ condensate with
angular momentum $J=1$ which locks to color
in such a way that rotational invariance is not
broken~\cite{Schaefer1Flavor}.
The resulting gap has been estimated to be of order
hundreds of keV~\cite{Schaefer1Flavor}, although applying
results of Ref.~\cite{BowersLOFF} suggests a somewhat smaller gap, around
10 keV.  The blue up and down quarks can also pair, forming
a $J=1$ condensate which breaks rotational invariance~\cite{ARW1}.
The related gap was estimated to be a few keV~\cite{ARW1}, but this 
estimate was not robust and should be revisited in light of more
recent developments given its importance
in the following.  The critical temperature $T_c$ above
which no condensate forms is of order the  zero-temperature gap
$\Delta$. ($T_c=0.57 \Delta$ for $J=0$ condensates~\cite{PisarskiRischke}.) 
Therefore, 
if there are quarks for which $\Delta\sim T$ or smaller, these quarks
do not pair at temperature $T$. Such quark quasiparticles 
will radiate neutrinos rapidly (via direct URCA
reactions like $d\rightarrow u+e+\bar\nu$, 
$u\rightarrow d+e^+ +\nu$, etc.)
and the quark matter core will cool rapidly and determine the
cooling history of the star as a whole~\cite{Schaab,Page}.
The star
will cool rapidly until its interior temperature is
$T<T_c\sim\Delta$, at which time the quark matter core will become
inert and the further cooling history will be dominated by
neutrino emission from the nuclear matter fraction of the star. 
If future data were to show that neutron
stars first cool rapidly (direct URCA) and then cool more
slowly, such data would allow an estimate of the smallest 
quark matter gap. We are unlikely to be so lucky.
The simple observation of rapid cooling would {\it not} be an unambiguous
discovery of quark matter with small gaps; there are other
circumstances in which the direct URCA processes occurs.
However, if as data on neutron star temperatures improves in coming
years the standard cooling scenario proves correct,
indicating the absence of the direct URCA processes, 
this {\it would} rule out the presence
of quark matter with gaps in the 10 keV range or smaller.  
The presence of a quark matter core
in which all gaps are $\gg T$ can never be revealed by
an analysis of the cooling history.

{\bf Supernova Neutrinos:}
We now turn from neutrino emission from a neutron star which
is many years old to that from the protoneutron star 
during the first seconds of  a supernova.
Carter and Reddy~\cite{CarterReddy}
have pointed out that when this protoneutron
star is heated up to its maximum temperature of order 30-50~MeV,
it may feature a quark matter core which is too hot for color
superconductivity.  As the core of the protoneutron star cools
over the coming seconds, if it contains quark matter this quark
matter will cool through $T_c$, entering the color superconducting
regime of the QCD phase diagram from above.  For $T\sim T_c$, the
specific heat rises and the cooling slows. Then, as $T$ drops
further and $\Delta$ increases to become greater than $T$,
the specific heat drops rapidly. Furthermore, as the number
density of quark quasiparticles becomes suppressed by $\exp(-\Delta/T)$,
the neutrino transport mean free path rapidly 
becomes very long~\cite{CarterReddy}.
This means that all the neutrinos previously trapped
in the now color superconducting
core are able to escape in a sudden burst.  If we 
are lucky enough that a terrestrial neutrino detector
sees thousands of neutrinos from a future supernova, Carter
and Reddy's results suggest that there may be a signature of the
transition to color superconductivity present in the time distribution
of these neutrinos.  Neutrinos from the core of the protoneutron
star will lose energy as they scatter on their way out, but because
they will be the last to reach the surface of last scattering, they
will be the final neutrinos received at the earth.  If they are emitted
from the quark matter core in a sudden burst, they may therefore
result in a bump at late times in the temporal distribution of
the detected neutrinos.  More detailed study remains to be done
in order to understand how Carter and Reddy's signature, dramatic
when the neutrinos escape from the core, is processed as the neutrinos
traverse the rest of the protoneutron star and reach their
surface of last scattering.

{\bf R-mode Instabilities:}  
Another arena in which color superconductivity comes into play
is the physics of r-mode instabilities.  A neutron star whose
angular rotation frequency $\Omega$ is large enough is unstable
to the growth of r-mode oscillations which radiate 
away angular momentum via gravitational waves, reducing $\Omega$.
What does ``large enough'' mean?  The answer depends on 
the damping mechanisms which act to prevent the growth of
the relevant modes.  Both shear viscosity and bulk viscosity
act to damp the r-modes, preventing them from going unstable.
The bulk viscosity and the quark contribution
to the shear viscosity both become exponentially
small in quark matter with $\Delta>T$ and as a result,
as Madsen~\cite{Madsen} has shown, 
a compact star made {\it entirely} of quark matter with
gaps $\Delta=1$~MeV or greater is 
unstable if its spin frequency is greater than tens to 100~Hz.
Many compact stars spin faster than this, and Madsen therefore
argues that compact stars cannot be strange quark stars
unless some quarks remain ungapped.  Alas, this powerful argument
becomes much less powerful in the context of a neutron star
with a quark matter core.  First, the r-mode oscillations 
have a wave form whose amplitude is largest at large radius,
outside the core.
Second, in an ordinary neutron star there
is a new source of damping: friction at the boundary between
the crust and the neutron superfluid ``mantle'' keeps the 
r-modes stable regardless of the properties of a quark matter 
core~\cite{Bildsten,Madsen}.

{\bf Magnetic Field Evolution:}
Next, we turn to the physics of magnetic fields within
color superconducting neutron star cores~\cite{Blaschkeflux,ABRflux}.  
The interior
of a conventional neutron star is a superfluid (because of neutron-neutron
pairing) and is an electromagnetic superconductor
(because of proton-proton pairing).  Ordinary magnetic fields
penetrate it only in the cores of magnetic flux tubes.
A color superconductor behaves differently. At first
glance, it seems that because a diquark Cooper pair has nonzero
electric charge, a diquark condensate
must exhibit the standard Meissner effect, expelling
ordinary magnetic fields or restricting them to flux tubes
within whose cores the condensate vanishes.  This is not
the case~\cite{ABRflux}.
In both the 2SC and CFL phase a linear combination
of the $U(1)$ gauge transformation of ordinary electromagnetism
and one (the eighth) color gauge transformation remains unbroken 
even in the presence of the condensate.  This means that 
the ordinary photon $A_\mu$ and the eighth gluon $G_\mu^8$
are replaced by new linear combinations
\begin{eqnarray}
A_\mu^{\tilde Q} &=& \cos\alpha_0 \,A_\mu + \sin\alpha_0\,G_\mu^8
\nonumber\\
A_\mu^{X} &=& -\sin\alpha_0\,A_\mu + \cos\alpha_0\,G_\mu^8
\end{eqnarray}
where $A_\mu^{\tilde Q}$ is massless and $A_\mu^{X}$ is massive.
That is,  $B_{\tilde Q}$ satisfies the ordinary Maxwell
equations while $B_X$ experiences a Meissner effect.
The mixing angle $\alpha_0$ is the analogue of the Weinberg
angle in electroweak theory, in which the 
presence of the Higgs condensate causes the $A_\mu^Y$ and the third
$SU(2)_W$ gauge boson to mix to form the photon, $A_\mu$, and 
the massive $Z$ boson.   
$\sin(\alpha_0)$ is proportional to $e/g$ and turns
out to be about $1/20$ in the 2SC phase and $1/40$ in the CFL
phase~\cite{ABRflux}.  This means that the 
$\tilde Q$-photon which propagates in color superconducting
quark matter is mostly photon with only 
a small gluon admixture. If a color superconducting neutron star core 
is subjected to an ordinary magnetic field, it will either
expel the $X$ component of the flux
or restrict it to flux tubes, but it can
(and does~\cite{ABRflux}) admit the great majority of the flux
in the form of a $B_{\tilde Q}$ magnetic field satisfying
Maxwell's equations.   
The decay in time of this ``free field'' (i.e. not in flux tubes) 
is limited by the $\tilde Q$-conductivity of the quark matter.
A color superconductor is not a $\tilde Q$-superconductor --- 
that is the whole point --- but it turns out 
to be a very good
$\tilde Q$-conductor due to the presence of electrons:
the $B_{\tilde Q}$ magnetic field decays only on a time scale 
which is much longer than the age of the universe~\cite{ABRflux}.
This means that a quark matter core within a neutron
star serves as an ``anchor'' for the magnetic field:
whereas in ordinary nuclear matter the magnetic flux
tubes can be dragged outward by the neutron superfluid
vortices as the star spins down~\cite{Srinivasan}, 
the magnetic flux within the 
color superconducting core simply cannot decay.
Even though this distinction is a qualitative one, it
will be 
difficult to confront it with data since what is
observed is the total dipole moment of the neutron star.
A color superconducting
core anchors those magnetic flux lines which pass through
the core, while in a neutron star with no quark matter core
the entire internal magnetic field can decay over time. 
In both cases, however, the total dipole moment can change
since the magnetic flux lines which do not pass through
the core can move.

{\bf Glitches in Quark Matter:}
The final consequence of color superconductivity 
we wish to discuss is the possibility that (some)
glitches may originate within quark matter regions of a 
compact star~\cite{BowersLOFF}.
In any context in which color superconductivity arises
in nature, it is likely to involve pairing between species of quarks
with differing chemical potentials.   
If the chemical potential difference  
is small enough, BCS pairing
occurs as we have been discussing. 
If the Fermi surfaces are too far apart, no pairing between the species is
possible. The transition between the BCS and unpaired states as the
splitting between Fermi momenta increases has been studied in
electron~\cite{Clogston} and QCD~\cite{ABR2+1,SW2,Bedaque}
superconductors, assuming that no other state intervenes.  However,
there is good reason to think that another state can occur.  This is
the ``LOFF'' state, first explored by Larkin and Ovchinnikov~\cite{LO}
and Fulde and Ferrell~\cite{FF} in the context of electron
superconductivity in the presence of magnetic impurities.
They found that near the
unpairing transition, 
it is favorable to form a state in
which the Cooper pairs have nonzero momentum. This is favored because
it gives rise to a region of phase space where each of the two quarks
in a pair can be close to its Fermi surface,
and such pairs can be created at low cost in free energy.
Condensates of this sort spontaneously
break translational and rotational invariance, leading to
gaps which vary periodically in a crystalline pattern.
If in some shell within the quark matter core
of a neutron star (or within a strange quark star)  
the quark number densities are
such that crystalline color superconductivity arises,
rotational vortices may be pinned in this shell, making
it a locus for glitch phenomena.

We~\cite{BowersLOFF} have explored the range of parameters
for which crystalline color superconductivity occurs in
the QCD phase diagram, upon making various simplifying assumptions.
For example, we focus primarily on a four-fermion interaction
with the quantum numbers of single gluon exchange.
Also, we only consider pairing between $u$ and $d$ quarks, with
$\mu_d=\bar\mu+\delta\mu$ and $\mu_u=\bar\mu-\delta\mu$, whereas
we expect a LOFF state when the difference between the Fermi momenta
of any two quark flavors is near an unpairing transition. 
We find the LOFF state is favored
for values of $\delta\mu$ which satisfy $\delta\mu_1 < \delta\mu 
< \delta\mu$
where $\delta\mu_1/\Delta_0=0.707$ and $\delta\mu_2/\Delta_0=0.754$ in the 
weak coupling limit in which $\Delta_0\ll \mu$. (Here, $\Delta_0$
is the 2SC gap that would arise if $\delta\mu$ were zero.)
The LOFF gap parameter decreases from $0.23 \Delta_0$
at $\delta\mu=\delta\mu_1$ (where there is a first order BCS-LOFF
phase transition)
to zero at $\delta\mu=\delta\mu_2$ (where there is a second order
LOFF-normal transition).  
Except for very close to $\delta\mu_2$, the critical
temperature above which the LOFF state melts will be much
higher than typical neutron star temperatures.
At stronger coupling the LOFF gap parameter decreases relative
to $\Delta_0$ and 
the window of $\delta\mu/\Delta_0$ within which the LOFF state
is favored shrinks. The window grows
if the interaction is changed to weight electric
gluon exchange more heavily than magnetic gluon exchange.

The quark matter which may be 
present within a compact star will be in
the crystalline color superconductor (LOFF) state 
if $\delta\mu/\Delta_0$ is in the requisite range.  
For a reasonable value of $\delta\mu$, say 25~MeV,
this
occurs if the gap $\Delta_0$ which characterizes the uniform
color superconductor present at smaller values of $\delta\mu$ is 
about 40~MeV. This is in the middle of the range of present
estimates.  Both $\delta\mu$ and $\Delta_0$ vary as a function
of density and hence as a function of radius in a compact star.
Although it is too early to make quantitative predictions,
the numbers are such that crystalline color superconducting
quark matter may very well occur in a range of radii within a compact 
star. It is therefore worthwhile to consider the consequences.

Many pulsars have been observed to glitch.  Glitches are sudden
jumps in rotation frequency $\Omega$ which may
be as large as $\Delta\Omega/\Omega\sim 10^{-6}$, but may also
be several orders of magnitude smaller. The frequency of observed
glitches is statistically consistent with the hypothesis that 
all radio pulsars experience glitches~\cite{AlparHo}.
Glitches are thought to originate from interactions
between the rigid crust, somewhat more than a kilometer thick in a typical
neutron star, and rotational vortices in the 
neutron superfluid which are moving (or trying to move) outward
as the star spins down. Although the models~\cite{GlitchModels} differ
in important respects, all agree that the fundamental requirements
are the presence of rotational vortices in a superfluid 
and the presence
of a rigid structure which impedes the motion of vortices and
which encompasses enough of the volume of the pulsar to contribute
significantly to the total moment of inertia.

Although it is 
premature to draw quantitative conclusions,
it is interesting to speculate that some glitches may originate 
deep within a pulsar which features
a quark matter core, in a region of that core 
in which the color superconducting quark matter is in
a LOFF crystalline color superconductor phase.
A three flavor analysis is required to determine whether the LOFF
phase is a superfluid.   If the only pairing is between $u$
and $d$ quarks, this 2SC phase is not a superfluid~\cite{ARW1,ABR2+1},
whereas if all three
quarks pair in some way, a superfluid {\it is} 
obtained~\cite{CFL,ABR2+1}.
Henceforth, we suppose  that the LOFF phase is a superfluid, 
which means that if it occurs within a pulsar it will be threaded
by an array of rotational vortices.
It is reasonable to expect that these vortices will
be pinned in a LOFF crystal, in which the
diquark condensate varies periodically in space.
Indeed, one of the suggestions for how to look for a LOFF phase in
terrestrial electron superconductors relies on the fact that
the pinning of magnetic flux tubes (which, like the rotational vortices
of interest to us, have normal cores)
is expected to be much stronger
in a LOFF phase than in a uniform BCS superconductor~\cite{Modler}.

A real calculation of the pinning force experienced by a vortex in a
crystalline color superconductor must await the determination of the
crystal structure of the LOFF phase. We can, however, attempt an order
of magnitude estimate along the same lines as that done by Anderson
and Itoh~\cite{AndersonItoh} for neutron vortices in the inner crust
of a neutron star. In that context, this estimate has since been made
quantitative~\cite{Alpar77,AAPS3,GlitchModels}.  
For one specific choice of parameters~\cite{BowersLOFF}, the LOFF phase
is favored over the normal phase by a free energy 
$F_{\rm LOFF}\sim 5 \times (10 {\rm ~MeV})^4$ 
and the spacing between nodes in the LOFF
crystal is $b=\pi/(2|{\bf q}|)\sim 9$ fm.
The thickness of a rotational vortex is
given by the correlation length $\xi\sim 1/\Delta \sim 25$ fm.  
The pinning energy
is the difference between the energy of a section of vortex of length 
$b$ which is centered on a node of the LOFF crystal vs. one which
is centered on a maximum of the LOFF crystal. It 
is of order $E_p \sim F_{\rm LOFF}\, b^3 \sim 4 {\rm \ MeV}$.
The resulting pinning force per unit length of vortex is of order
$f_p \sim E_p/b^2 \sim  (4 {\rm \ MeV})/(80 {\rm \ fm}^2)$.
A complete calculation will be challenging because
$b<\xi$, and is likely to yield an $f_p$
which is somewhat less than that we have obtained by dimensional 
analysis.
Note that our estimate of $f_p$ is
quite uncertain both because it is
only based on dimensional analysis and because the values
of $\Delta$, $b$ and $F_{\rm LOFF}$ are 
uncertain.  (We hae a good understanding of 
all the ratios $\Delta/\Delta_0$, $\delta\mu/\Delta_0$, $q/\Delta_0$ 
and consequently $b\Delta_0$ in the LOFF phase.  It is 
of course the value of the BCS gap $\Delta_0$ which is uncertain.) 
It is premature to compare our crude result 
to the results of serious calculations 
of the pinning of crustal neutron vortices as in 
Refs.~\cite{Alpar77,AAPS3,GlitchModels}.  It is nevertheless
remarkable that they prove to be similar: the pinning
energy of neutron vortices in the inner crust 
is $E_p \approx 1-3  {\rm \ MeV}$
and the pinning force per unit length is
$f_p\approx(1-3 {\rm ~MeV})/(200-400 {\rm ~fm}^2)$.
Perhaps, therefore, glitches occurring in a region of crystalline
color superconducting quark matter may yield similar phenomenology
to those occurring in the inner crust.

Perhaps the most interesting consequence of these speculations
arises in the context of compact stars made entirely of 
strange quark matter.  The work of Witten~\cite{Witten}
and Farhi and Jaffe~\cite{FarhiJaffe} raised the possibility
that strange quark matter may be energetically stable relative
to nuclear matter even at zero 
pressure.
If this is the
case it raises the question whether observed compact stars---pulsars,
for example---are strange quark stars~\cite{HZS,AFO} rather than
neutron stars.  
A conventional neutron star may feature
a core made of strange quark matter, as we have been discussing 
above.\footnote{Note that a convincing discovery of
a quark matter core within an otherwise hadronic
neutron star would demonstrate
conclusively that strange quark matter is {\it not} stable
at zero pressure, thus ruling out the existence of strange
quark stars.  It is not possible for 
neutron stars with quark matter cores and strange quark
stars to both be stable.}
Strange quark stars, on the other hand, are made (almost)
entirely of quark
matter with either no hadronic matter content at all or
with a thin crust, of order one hundred meters thick, which contains
no neutron superfluid~\cite{AFO,GlendenningWeber}.
The nuclei in this thin crust
are supported above the quark matter by electrostatic forces;
these forces cannot support a neutron fluid.  Because
of the absence of superfluid neutrons, and because of the thinness of
the crust, no successful models of glitches in the crust
of a strange quark star have been proposed.  
Since pulsars are observed to glitch, the apparent lack of a 
glitch mechanism for strange
quark stars  has been the 
strongest argument that pulsars cannot be strange quark 
stars~\cite{Alpar,OldMadsen,Caldwell}.
This conclusion must now be revisited.  

Madsen's conclusion~\cite{Madsen} that a strange
quark star is prone to r-mode instability due to
the absence of damping must
also be revisited, since the relevant fluid oscillations
may be damped within or at the boundary of a region
of crystalline color superconductor.

The quark 
matter in a strange quark star, should
one exist, would be a color superconductor.
Depending on the mass of the star, the 
quark number densities increase by a factor of about two to ten
in going from the surface to the center~\cite{AFO}. This means
that the chemical potential differences among the three
quarks will vary also, and there could be a range of radii
within which the quark matter is in a crystalline
color superconductor phase.  This raises the 
possibility of glitches in strange quark stars.
Because the
variation in density with radius is gradual, if a shell
of LOFF quark matter exists it need not be particularly thin.
And, we have seen, the pinning forces may be comparable
in magnitude to those in the inner crust of a conventional
neutron star.
It has recently been suggested (for reasons unrelated to our considerations)
that certain accreting compact stars
may be strange quark stars~\cite{Bombaci}, although the
evidence is far from unambiguous~\cite{ChakrabartyPsaltis}.
In contrast, 
it has been thought that, because they glitch,  
conventional radio pulsars cannot be strange
quark stars.  Our work questions this assertion
by raising the possibility that glitches
may originate within a layer of quark matter 
which is in a crystalline color superconducting state.

{\bf Closing Remarks:}
The answer to the question of whether the QCD phase diagram 
does or does not feature a 2SC interlude
on the horizontal axis, separating the CFL and baryonic phases
in both of which chiral symmetry is broken,
depends on whether the strange
quark is effectively heavy or effectively light.  This is the
central outstanding qualitative question about 
the high density region of the QCD phase diagram.
A central question at higher temperatures, 
namely where does nature locate
the critical point $E$, also depends on the strange quark mass. 
Both questions are hard to answer theoretically with
any confidence.  The high temperature region 
is in better shape, however, because the program of experimentation
described in Section II allows 
heavy ion collision experiments to search for 
the critical point $E$. 
Theorists have described how to use phenomena characteristic of freezeout
in its vicinity to discover $E$;  this gives experimentalists
the ability to locate it convincingly.
The discovery of $E$ would allow us to draw the higher temperature
regions of the map of the QCD phase diagram in ink.
At high density, there has been much recent progress in
our understanding 
of how the presence of color superconducting quark matter
in a compact star would affect five different phenomena:
cooling by neutrino emission, the temporal pattern of the
neutrinos emitted by a supernova, the evolution of neutron
star magnetic fields, r-mode instabilities, and glitches.
Nevertheless, much theoretical work remains to be done before
we can make sharp proposals for which 
astrophysical observations are most likely to help teach us how to
ink in the boundaries of the 2SC and CFL
regions in the QCD phase diagram.  
Best of all, though, and as in heavy ion physics, a wealth of new
data is expected over the next few years.

\end{document}